# Direct Evidence of Superconductivity with Twofold Symmetry in $Bi_2Te_3$ Thin Film Deposited on $FeTe_{0.55}Se_{0.45}$

Minyang Chen,[1] Xiaoyu Chen,[1] Huan Yang,[1,2]* Zengyi Du,[1] Hai-Hu Wen[1,2]*

[1] National Laboratory of Solid State Microstructures and Department of Physics, Nanjing University, Nanjing 210093, China

[2] Collaborative Innovation Center of Advanced Microstructures, Nanjing University, Nanjing, 210093, China

*Corresponding author. Email: huanyang@nju.edu.cn (H. Y.); hhwen@nju.edu.cn (H.H.W.)

**Abstract**: **Topological superconductor is a timely and frontier topic in condensed matter physics. In superconducting state, an order parameter will be established with the basic or subsidiary symmetry of the crystalline lattice. In doped $Bi_2Se_3$ with a basic threefold symmetry, it was predicted however that superconductivity may have a twofold symmetry of odd parity. Here we report the proximity effect induced superconductivity in $Bi_2Te_3$ thin film on top of an iron-based superconductor $FeTe_{0.55}Se_{0.45}$. By using the quasiparticle interference technique, we demonstrate clear evidence of twofold symmetry of the superconducting gap. The gap minimum is along one of the main crystalline axis following the so-called $\Delta_{4y}$ notation. This is also accompanied by the elongated vortex shape mapped out by the density of states within the superconducting gap, with probably the Majorana mode within the vortex core. Our results reveal the direct evidence of superconductivity with odd parity in $Bi_2Te_3$ thin film.**

**One Sentence Summary:** Superconductivity with twofold symmetry is observed by scanning tunneling microscopy in $Bi_2Te_3$ thin film, yielding the direct evidence of odd parity.

Superconductivity with order parameters of odd parity is highly desired because of the expectation of Majorana Fermions and potential applications in quantum computation (*1,2*). The topological superconductors are supposed to have order parameters with odd parity, which has become a hot topic in nowadays condensed matter physics. One of the signatures of odd parity is the existence of a superconducting (SC) order parameter with a twofold symmetry, such as $p_x$- or $p_y$-wave pairing. But in reality it is extremely rare. In some compounds with self-doping, such as $M_xBi_2Se_3$ (M = Cu, Nb and Sr), theoretically it was predicted that two dimensional topological SC state (*3,4*) may occur. Some traces of this long sought topological superconductivity have been reported together with the observation of the signatures of Majorana fermions or modes (*5,6*). Proximity effect may be another way to induce possible topological superconductivity (*7*), which has been well demonstrated in $Bi_2Se_3$ thin film on top of 2*H*-$NbSe_2$ (*8*) and with the possible evidence of Majorana mode within the vortex core on $Bi_2Te_3$/$NbSe_2$ heterostructure (*9*). In the doped materials $M_xBi_2Se_3$ (M = Cu, Nb and Sr), bulk superconductivity were observed (*10-12*) and full gaps have been reported on the surface through scanning tunneling microscopy/spectroscopy (STM/STS) measurements (*13,14*). Recently, Knight-Shift of NMR measurements (*15*) and angle-resolved specific heat (*16*) reveal a twofold feature of superconductivity when the magnetic field rotates within the *ab*-plane. Theoretically Fu (*17*) did a quantitative assessment by taking account the spin orbital coupling and the multi-orbital effect in $M_xBi_2Se_3$ and predicted that it could be possible to have a fully gapped order parameter but with odd parity, namely a twofold symmetry. Beside the NMR and specific heat measurements (*15,16*), the *c*-axis resistivity also exhibits a twofold symmetry when the applied magnetic field is rotated within the basal plane in $Sr_xBi_2Se_3$ (*18,19*). However, to extract the information of gap structure from these data we inevitably need some models to relate the gap minimum with the crystalline lattice. *We are lacking of a direct evidence for this very unique pairing order parameter with odd parity.* In this work, based on the STM/STS measurements on superconductivity induced by proximity effect in $Bi_2Te_3$ thin film on top of the iron-based superconductor $FeTe_{0.55}Se_{0.45}$ we present the clear evidence of a twofold symmetry of the SC order parameter and pin down the gap form of $\Delta_{4y}$ from several candidates (*16,17*). We also observe elongated vortex structure associated with this effect.

Single crystals of $FeTe_{0.55}Se_{0.45}$ with the transition temperature $T_c$ of about 13.3 K are used as the substrates. After cleaving the $FeTe_{0.55}Se_{0.45}$ crystal in a high vacuum, we obtain a fresh and clean surface on which we grow the topological insulator (TI) $Bi_2Te_3$ film (20) by using the molecular beam epitaxy (MBE) technique. As shown in Fig. 1A, the surface topography of $FeTe_{0.55}Se_{0.45}$ shows a structure of square lattice. The white and dark atoms were supposed to be the Te and Se atoms respectively (21). The atomic-resolved topographic image of the resultant $Bi_2Te_3$ film is shown in Fig. 1B, and it is clear that the top Te-terminal surface has a hexagonal structure. With the deposition rate of about 0.5 QL/min, we obtain the $Bi_2Te_3$ thin film with very good connectivity. The thickness changes from place to place, a typical case is shown in Fig. 1C. The height difference between two sides of a step is about 1 nm, which corresponds to the thickness of a single QL of $Bi_2Te_3$. The exact thickness of $Bi_2Te_3$ film in local area is determined by the tunneling spectra (9) with wide-range of bias voltage, which is shown in Fig. 1D. The general shape of the spectrum measured on the $Bi_2Te_3$ film with the local thickness larger than 1 QL is similar to that measured on $Bi_2Te_3/NbSe_2$ heterostructure (9), as indicated by the arrow reflecting the kinky position. However, it should be noted that the features of conduction band minimum seem to be absent on the spectra, and the energy position of the kink for the 1 QL film is very different from that for 1 QL $Bi_2Te_3/NbSe_2$ heterostructure. In regions with different thickness, superconductivity is successfully induced by proximity effect, which can be observed on the tunneling spectra in lower-energy range, and the typical spectra are shown in Fig. 1E. The spectra measured on the $FeTe_{0.55}Se_{0.45}$ substrate show a full gapped feature with some non-uniform shape, which was described in our previous work (22). The typical feature of the spectra on the $FeTe_{0.55}Se_{0.45}$ substrate is that one or two pairs of coherence peaks can be observed with peak energies varying from 1.1 to 2.1 meV. As shown in Fig. 1E, the tunneling spectrum measured on 1 QL $Bi_2Te_3$ also has the trace of such feature of coherent peaks of $FeTe_{0.55}Se_{0.45}$. However, the homogeneity of tunneling spectra becomes much better on $Bi_2Te_3$ films with thickness larger than 2 QLs, which is presented fig. S1. Meanwhile, with increasing the thickness of $Bi_2Te_3$ thin film, the energy difference between two coherent peaks shrinks, and the density of states (DOS) at zero-bias also become higher. This indicates that the superconductivity in the $Bi_2Te_3$ thin film is induced by proximity effect. Interestingly, on the surfaces with the thickness of and larger than 2 QL, the spectra show a V-shape near the Fermi energy. The general shape and the evolution of the spectra with thickness are similar to previous results on $Bi_2Te_3/NbSe_2$ heterostructure (9).

Since the proximity effect induced superconductivity has been well formed in the 2 QL $Bi_2Te_3/FeTe_{0.55}Se_{0.45}$ heterostructure, it is worth to detect the pairing symmetry of superconductivity. The quasiparticle interference (QPI) technology is a useful tool to detect the SC gap structure (*23,24*), hence we do the measurement on the 2 QL Bi$_2$Te$_3$/FeTe$_{0.55}$Se$_{0.45}$ heterostructure in order to see whether we have the twofold SC gap. The standing waves of the electrons for the QPI measurements, which contains information in *k*-space, are scattered by impurities and they interfere each other forming a certain pattern in space. In our sample, generally it is difficult to find a place with enough impurities because of the good quality of the film. However, successful QPI measurements were carried out in places with some Te vacancies, and the topography of a typical case is shown in fig. S2A. Figure 2, A to J, show the corresponding Fourier transformed (FT) real-space QPI patterns (fig. S2) in *q*-space. When the bias voltage reaches the SC gap energy of about 2.0 meV and above, the FT-QPI pattern shows a frame of hexagon. The FT-QPI at 2.0 mV shows very clear feature of sixfold symmetry, which is consistent with the auto-correlation simulation result from a 2D hexagonal Fermi surface (fig. S3). We refer this sixfold FT-QPI pattern measured at ±2 mV and above as the normal state one, which is reasonable by considering the crystal structure of Bi$_2$Te$_3$. The stronger intensity shows up at the position near every Γ-K direction, while the intensity near the Γ-M directions is much weaker, this is consistent with the self-correlation simulation (see fig. S3). If the measuring bias voltage is below 0.5 mV, no clear intensity can be observed along the hexagon from the FT-QPI comparing with that at 2.0 mV. It suggests that the SC gap is nodeless, and the gap minimum, if existing, is larger than 0.5 meV. When the bias voltage is between 1.0 to 1.5 mV, one can find that the intensity of FT-QPI first emerges along one pair of Γ-K directions marked by double arrowed green line in Fig. 2A. This feature can also get support from the real-space QPI images (fig. S2, B to J) as some parallel standing waves are clearly observed. The same feature appears at the negative bias of −1.0 to −1.5 mV. When the bias voltage exceeds 1.5 mV, the other four sides of the hexagon show up. The hexagon becomes complete when the bias voltage is about ±2.5 mV (not shown here). Based on the Bogoliubov theory of DOS in the superconducting state, a reasonable explanation for the evolution of FT-QPI pattern is a twofold superconducting gap with gap minima along one pair of Γ-K direction. The twofold symmetry feature is better illustrated in the angular dependence of averaged intensity per pixel, as shown in Fig. 2K. The intensity peaks near one pair of opposite Γ-K directions ($\phi = 60°$ and $240°$) obviously break the rotation threefold symmetry of the crystal structure. If we compare the result with the Te-terminal surface, the gap minima locate along one of the crystallographic axis.

We also tried to use the twofold SC gap functions $\Delta = \Delta_{max}[(1-x)\cos 2\phi + x]$ to fit the tunneling spectra, and the fitting results are shown in fig. S4. Obviously the models with only one twofold SC gap can fit the spectra well when they are measured on $Bi_2Te_3$ with thickness larger than 2 QLs. The spectrum measured on 1 QL $Bi_2Te_3/FeTe_{0.55}Se_{0.45}$ heterostructure (fig. S4A) should be fitted by using the model two anisotropic gaps, and this may suggest that the spectrum measured on 1 QL $Bi_2Te_3/FeTe_{0.55}Se_{0.45}$ carries still some feature from the substrate $FeTe_{0.55}Se_{0.45}$. To illustrate the gap structure from our FT-QPI measurements, in Fig. 2M, we plot the resultant angle dependent SC gap for the 2 QL $Bi_2Te_3$ in the same coordinate plane of the schematic contour of the Fermi surface. The gap structure shows a twofold symmetry in *k*-space with minimum in one of the Γ-K directions. The twofold paring symmetry has been studied theoretically (*17*), and the direction of the gap minimum found here is consistent with one of the predicted results. According to our experimental results, the notation $\Delta_{4y}$ (Fig. 2N) is the only correct paring symmetry among all the proposals raised in Ref. 17. This set of experiments have been repeated in three rounds and all lead to the same observation, without exception.

The structure of a single vortex in a type-II superconductor usually gives rise to the information of the SC gaps (*25*). For example, we can obtain the SC gap Δ by the coherence length extracted from the STM image via $\xi = \hbar v_F/\pi\Delta$, and here $v_F$ is the Fermi velocity. Since the tunneling spectra look very similar to the ones on $Bi_2Te_3/NbSe_2$ heterostructure when the film is thicker than 2 QLs, it is quite curious to know whether we have the elongated vortex core due to this odd-parity superconductivity with twofold symmetry. We thus measure the vortex image on the 2 QL $Bi_2Te_3/FeTe_{0.55}Se_{0.45}$ heterostructure and show the main results in Fig. 3. One can see that the vortices are elongated along one principal crystalline axis, namely *a*-axis here (Fig. 3, A and B), which is very different from the vortex measured on the bare $FeTe_{0.55}Se_{0.45}$ crystal without $Bi_2Te_3$ (fig. S5B) (*22*). By a fitting to the spatial evolution of the zero-bias differential conductance with a model of exponential decay we obtain the anisotropy γ of the vortex, which is about 2.6. The fitting results are given in section I of (*20*) and shown in fig. S6. Intuitively, the elongated vortices reflect the in-plane anisotropy of coherence length, which is resulted from the anisotropic SC gap structure in *k*-space. The gap-minimum direction, which is determined simply from the longer coherence length, is consistent with the conclusion from the QPI measurement, i.e., along crystalline *a*-axis of the sample. Such elongated shape of vortices can be observed in superconductors with twofold symmetric crystal structure (*26,27*). However, for $Bi_2Te_3$, the Fermi velocity is roughly isotropic (*28,29*), and obviously it is difficult to understand such elongated vortices in a system with a threefold symmetry.

We also investigate the vortex bound states at the center of the vortex core in these elongated vortices on the 2 QL $Bi_2Te_3$/$FeTe_{0.55}Se_{0.45}$ heterostructure. The anisotropic spatial evolution of the CdGM state can reflect the information of the SC gap (*30*). The spatial evolutions of vortex core state along and perpendicular to the *a*-axis are shown in Fig. 3, C and D. One can find very different vortex bound state features when the STM tip moves crossing the elongated vortex. The spectra along the direction with short vortex core size shows a central peak at the vortex center, then it evolves into two symmetric peaks when moving away from the center. This is very similar to the case in a conventional *s*-wave superconductor with large $E_F$ (*31*). However, we must mention that, this feature is already very different from the case in the bare $FeTe_{0.55}Se_{0.45}$ crystals in which we observed a round vortex core with the discrete energy levels of Caroli-de Gennes-Matricon (CdGM) states due to the small Fermi energy (*22*). Thus we can conclude that the vortex core states on the 2 QL $Bi_2Te_3$/$FeTe_{0.55}Se_{0.45}$ heterostructure are very different from that of $FeTe_{0.55}Se_{0.45}$ due to very different values of Fermi energies. We need to emphasize that the superconductivity observed in $Bi_2Te_3$/$FeTe_{0.55}Se_{0.45}$ heterostructure here is intrinsic although it comes from the proximity effect. If we go along the elongated direction of the vortex, namely the *a*-axis direction, the evolution of spectra looks very spectacular. *A central peak appears at the core center, stays at zero energy and does not split at all when moving away from the center.* At this moment we still don't have a definite explanation of these contrasting behaviors along the two perpendicular directions. However, combining the evidence of odd parity superconductivity with two fold symmetry shown above, we would propose that this frozen zero-bias conductance peak along the elongated vortex direction may be induced by the complex mixture of a Majorana mode with the CdGM states. Actually a small bound state peak appears and remains at zero energy on the spectra measured in the direction of short vortex core size together with the two CdGM splitting peaks away from the vortex center, as marked by a mauve arrow in Fig. 3D. This interesting picture is highly desired for theoretical efforts. The elongated vortex image along *a*-axis strongly suggest that the gap minima presents along one of the crystal lattice directions. We also repeat the measurements at different places and different samples (fig. S7), elongated vortices are always observed.

In the following we demonstrate the twofold superconductivity via measuring the vortex image at different energies. As shown above, the vortices have elongated shapes as evidenced by the measurements of the zero-bias conductance mapping. It is very interesting to see how the vortex image evolves with the measuring bias energy if the superconductor has indeed a very anisotropic SC gap. We try to image the vortices at different energies in the area whose topography is shown in Fig. 4A. One can find that there is an edge-like dislocation in the upper right corner which comes probably from the substrate. The elongated vortices can be observed at zero-bias mapping, as shown in Fig. 4B. The elongated direction is again along the *a*-axis and has no angular relationship with the direction of the edge-like dislocation. Thus the elongated vortices or twofold superconducting gaps are intrinsic properties of the sample and not the effect of external causes. If the measuring bias voltage is increased but still below the maximum gap of about 2 meV, the two fold image feature appears still, but the vortices change their single elongated shape to pairs of parallel-elongated patterns, as shown here by Fig. 4, C to F, which may be originated from the different spatial evolution of the CdGM bound state peaks along two directions. We are aware of that, by increasing the measuring bias voltage from zero to a subgap value in LiFeAs, the vortex image changes from a bright spot to a hollow square with fourfold symmetry (*32*), which may share the same reason as our present case. When the bias voltage is 2.5 mV and above the SC gap maximum, as shown here by Fig. 4, G and H, the vortex image looks like a roughly isotropic dark disk without a twofold shape. The dark disk feature of the vortices measured at 2.5 mV can be understood by the relative reduction of DOS within a certain region in the center compared with that outside that region. When the energy is larger than the gap maximum, the spatial distribution of DOS along the two characteristic directions does not show clear difference, hence it is reasonable that we find such roughly isotropic image of vortices. This is a proof that the superconducting gap has the twofold symmetry. This interesting evolution of vortex images clearly warrants further theoretical investigations. The elongated vortices have been observed in all designed experiments on the same and different samples, without exception.

By doing STM/STS measurements and detailed QPI analysis on the 2 QL $Bi_2Te_3$/$FeTe_{0.55}Se_{0.45}$ heterostructure, we observe the twofold SC gap with the gap minima along one of the principal crystal axes, which is well consistent with the theoretical prediction (*17*). We regard this as a direct observation of the twofold symmetry of superconductivity in the related systems. Since our measurements are done in areas of microscopic scale with atomic resolution, this naturally removes the concern of possible inhomogeneity in bulk samples (*15,16*). According to the theory, this twofold symmetry is a natural consequence of the pairing with odd parity of the

$\Delta_{4y}$ notation. Thus it is very helpful to understand the properties of the possible topological superconductivity in the system. One interesting issue is that, why such twofold symmetry of superconductivity and the elongated vortices haven't been observed on the $Bi_2Te_3$/$NbSe_2$ heterostructure (9). We argue that this may be attributed to the large scale of penetration depth and coherence length of $2H$-$NbSe_2$ which makes the vortex shape on the substrate transmitted to the top $Bi_2Te_3$ surface. For the $Bi_2Te_3$/$FeTe_{0.55}Se_{0.45}$ heterostructure in our experiment, the influence from the substrate may be greatly suppressed because of the short coherence length in $FeTe_{0.55}Se_{0.45}$, and the vortices have their intrinsic shape which reflects its SC order parameter. The freezing of the central peak of vortex core states along the elongated direction when moving away from the vortex center may suggest the mixture of a Majorana mode and the CdGM states, which is certainly a very interesting proposal and needs further efforts to verify. Our experimental results bring about deep insight to the understanding of topological superconductivity in this system.

**Acknowledgments:** We acknowledge the useful discussions with Qianghua Wang. This work was supported by National Key R&D Program of China (grant no. 2016YFA0300400), National Natural Science Foundation of China (grant nos. 11534005, 11374144, and Natural Science Foundation of Jiangsu (grant no. BK20140015).

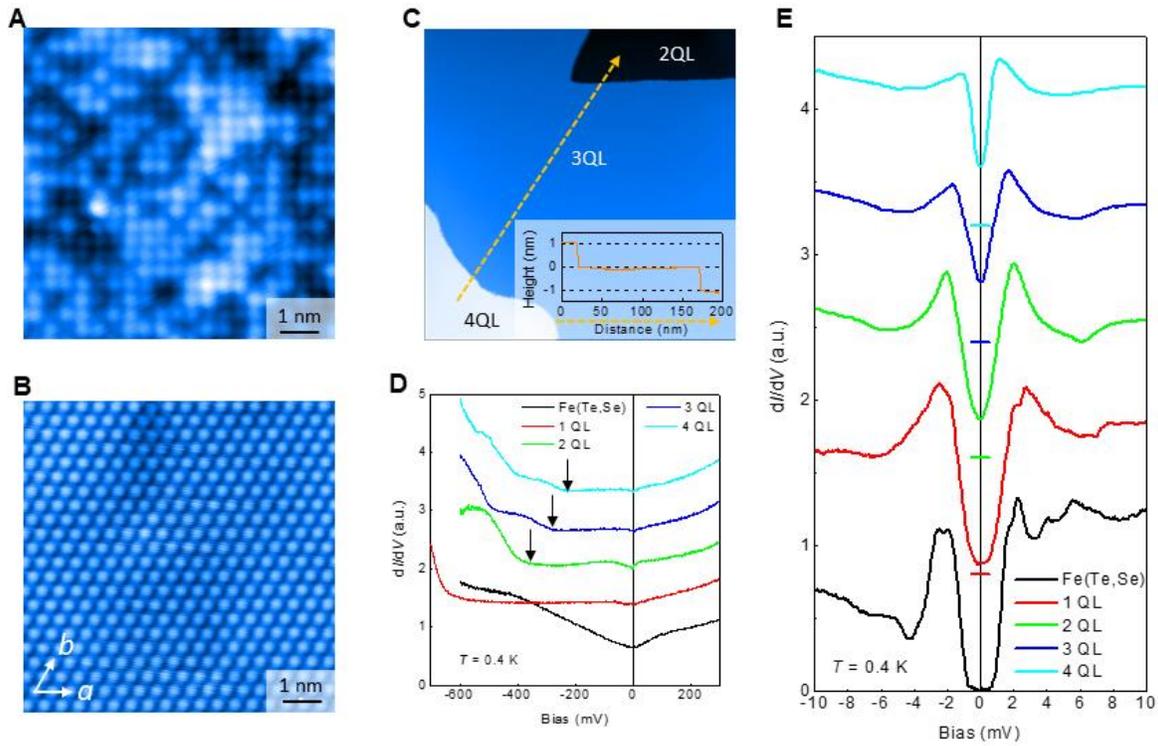

**Fig. 1**. STM/STS characterization of $Bi_2Te_3/FeTe_{0.55}Se_{0.45}$ heterostructure. (**A** and **B**) Atomically resolved topography (with bias voltage $V_{bias}$ = 10 mV, and tunneling current $I_t$ = 200 pA) of the (**A**) $FeTe_{0.55}Se_{0.45}$ substrate and (**B**) $Bi_2Te_3$ thin film. The measured lattice constants of the top Te/Se surface on $FeTe_{0.55}Se_{0.45}$ with square lattice and Te surface on $Bi_2Te_3$ with hexagonal lattice are 3.80 and 4.38 Å, respectively. (**C**) Topographic image ($V_{bias}$ = 2 V, $I_t$ = 10 pA, 150 × 150 nm$^2$) of $Bi_2Te_3$ film with different thickness. The height difference for each step along the dashed yellow arrowed line is about 1.0 nm which equals the thickness of 1 QL. (**D**) Typical differential conductance spectra (set-point bias voltage $V_{set}$ = 0.45 V, $I_{set}$ = 300 pA) measured on $FeTe_{0.55}Se_{0.45}$ substrate and $Bi_2Te_3$ thin films with different thickness. The black arrows show the positions of the typical kink features probably from the bulk valence band maximum or the Dirac point of the $Bi_2Te_3$ films of different thickness. (**E**) Tunneling spectra with superconducting feature ($V_{set}$ = 100 mV, $I_{set}$ = 50 pA) measured on $FeTe_{0.55}Se_{0.45}$ substrate and 1 to 4 QL $Bi_2Te_3/FeTe_{0.55}Se_{0.45}$. The short horizontal bars with the same color mark the zero differential conductance for each curve.

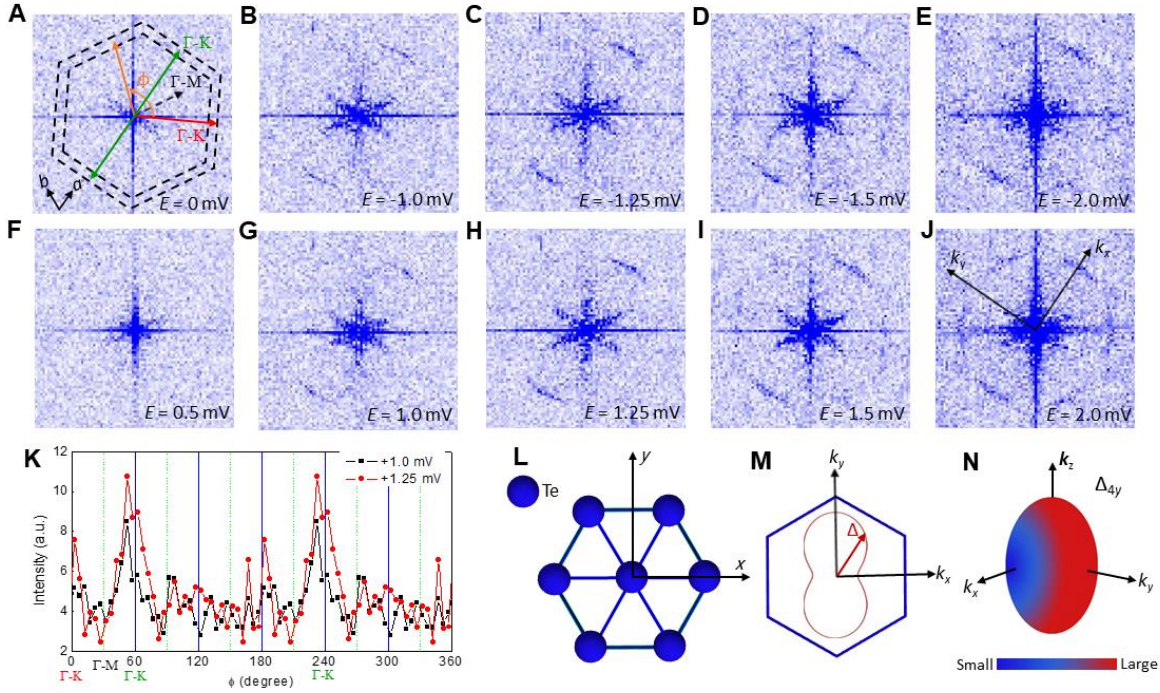

**Fig. 2.** Twofold superconducting gap resolved by QPI measurements. (**A** to **J**) The FT-QPI images derived from Fourier transformation to the QPI images with the real-space area of 84 × 84 nm$^2$ ($V_{set}$ = 10 mV, $I_{set}$ = 50 pA). (**K**) The averaged FT-QPI intensity per pixel in the $q$-space within the band bounded by the two parallel dashed hexagons in **A** with increment of every 5 degrees. The initial angle of ϕ is defined from one Γ-K direction as marked by a red arrow in **A**. (**L**) Schematic top view of the Te atom layer of Bi$_2$Te$_3$ surface. (**M**) Schematic 2D hexagonal Fermi surface and the resultant angular dependence of the anisotropic SC gap. (**N**) Schematic image of SC gap structures with an oval Fermi surface, the color gives a qualitative distribution of the SC gap.

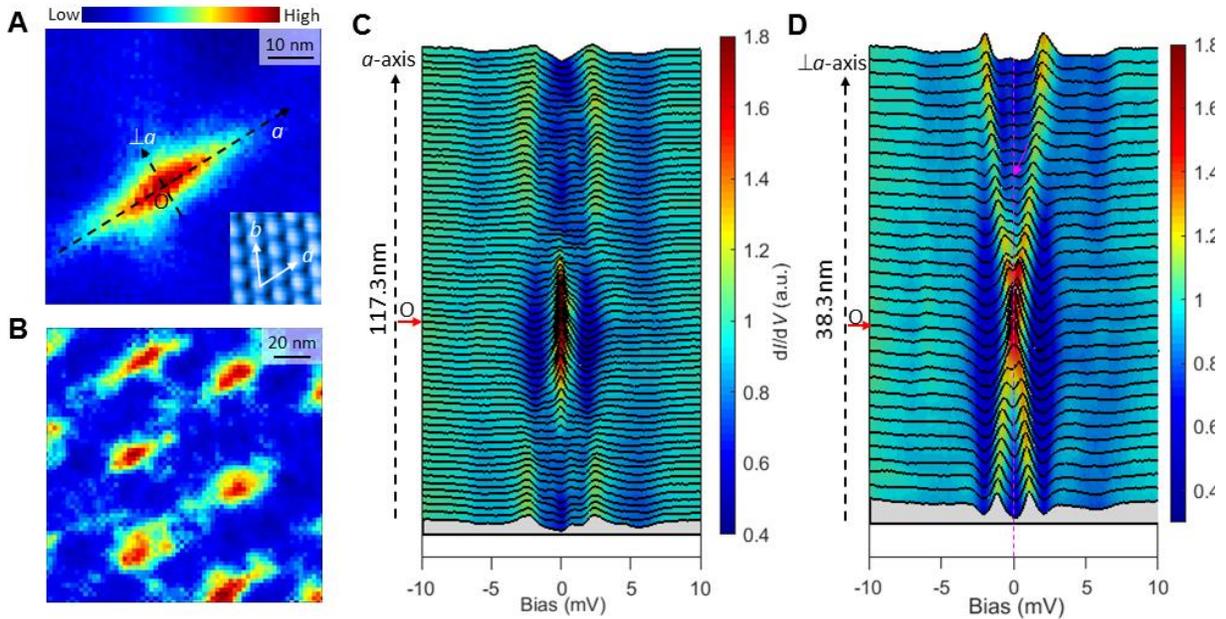

**Fig. 3.** Vortex image and vortex core states. (**A**) Zero-bias differential conductance map of a single vortex. The inset shows the topography measured near the vortex, and the vortex is elongated along $a$-axis. (**B**) Spectroscopic image of the vortex lattice consisted by elongated vortices measured at zero bias. (**C** and **D**) Tunneling spectra measured along $a$-axis direction and perpendicular to $a$-axis, respectively. The curves marked by red arrow represent the spectrum measured at the center of the vortex core. It should be noted that the dashed lines in **A** only denote the directions of the spectra measurements, but do not represent the spatial distance of the measurements. The spectra shown in **C** and **D** are taken along much longer lines than those shown in A, and the real spatial distance for measuring spectra are given by the vertical dashed arrowed line on the left-hand side of each panel. The measurements of all data in this figure are taken at $T = 0.4$ K and $B = 0.7$ T with $V_{set} = 10$ mV and $I_{set} = 50$ pA.

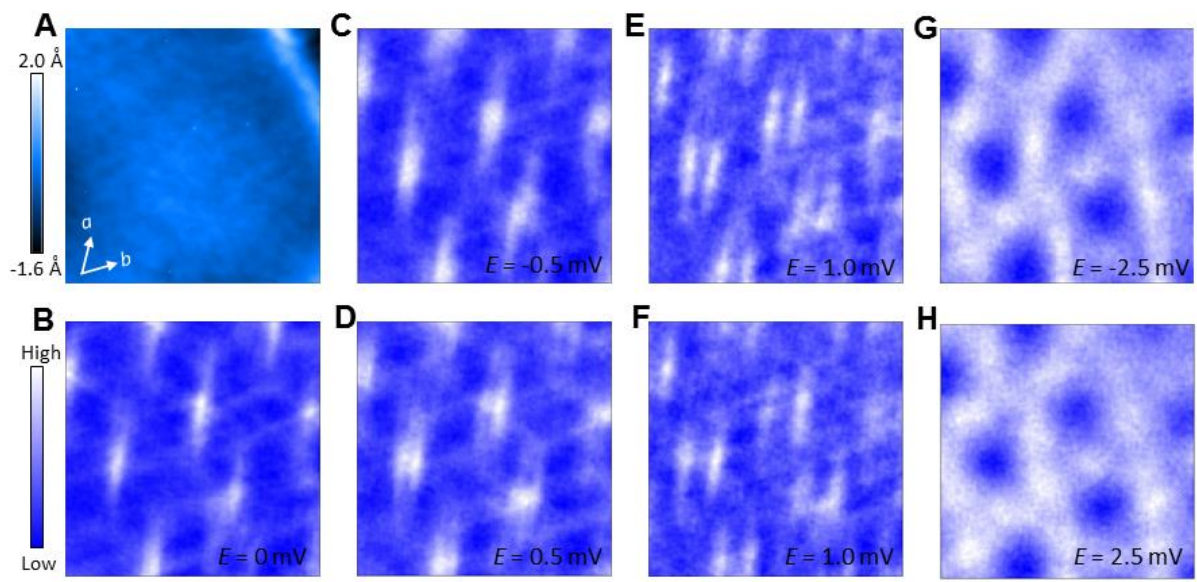

**Fig. 4.** Energy evolution of vortex image. (**A**) Topography of the view ($V_{bias}$ = 1 V, $I_t$ = 10 pA, 140 × 140 nm$^2$). (**B** to **H**) Vortex image measured with different bias voltage ($V_{set}$ = 10 mV, $I_{set}$ = 50 pA) at 0.4 K and 0.7 T.

**Supplementary Materials:**
Materials and Methods
Figures S1-S7
Tables S1
References (*33-36*)

## Supplementary Materials:

### Materials and Methods

The FeTe$_{1-x}$Se$_x$ (with nominal composition of $x = 0.45$) single crystals were grown by self-flux method (*33*). The excessive interstitial Fe atoms were eliminated by annealing the sample at 400 °C for 20 hours in O$_2$ atmosphere followed by quenching in the liquid nitrogen. The FeTe$_{0.55}$Se$_{0.45}$ samples were cleaved in an ultra-high vacuum with a base pressure about $1\times10^{-10}$ torr in the STM chamber. Then the tunneling spectra were measured before the film growth.

High-quality Bi$_2$Te$_3$ films were grown layer by layer on the FeTe$_{0.55}$Se$_{0.45}$ substrates in a Unisoku molecular beam epitaxy system (MBE) chamber which is connected directly to the STM operation chamber. High-purity Te (99.999%) and Bi (99.999%) were evaporated from the Createc effusion cells simultaneously with the flux ratio of about 18:1. The substrate temperature is about 265 °C during the film growth, and the film growth rate is about 0.5 QL per minute. The process of film growth was monitored by the technique of reflection high-energy electron diffraction (RHEED) in the MBE chamber.

The STM/STS measurements were carried out in a scanning tunneling microscope (USM-1300, Unisoku Co., Ltd.) with ultra-high vacuum, low temperature and high magnetic field. Tungsten tips were used for all the STM/STS measurements. A typical lock-in technique was used for the tunneling spectrum measurements with an AC modulation of 0.3 mV and 973.8 Hz. The tunneling spectra were recorded with fixed tip height with some set-point tunneling conditions, and then the *I-V* curves and d*I*/d*V* - *V* spectra were measured with such fixed tip height.

### Supplementary Text

### I. Calculation of penetration depth along two perpendicular directions of the elongated vortex

The spatial evolutions of differential conductance along and perpendicular to *a*-axis direction in Fig. 3A are shown in fig. S6, and the center of the vortex is set as the origin point. Then we fit the differential conductance $G(r)$ data by an exponential decay law (*34,35*) as $G(r) = \exp[-\mathrm{abs}(r)/\xi]+G(\infty)$, where the coherence length $\xi$ and differential conductance far away from the vortex $G(\infty)$ are fitting parameters. The fitting results are also shown in fig. S6, and the resultant coherence length $\xi_{\perp a}$ = 12.7 nm, $\xi_{\parallel a}$ = 33.5 nm. Hence, the anisotropy of coherence length $\gamma = \xi_{\parallel a}/\xi_{\perp a}$ calculated from the vortex is about 2.6.

## II. Dynes model fitting to the tunneling spectra measured on the Bi$_2$Te$_3$ films of different thickness

We fit the tunneling spectra by Dynes model (*36*) with a single anisotropic SC gap $\Delta(\phi)$ via

$$I = \int_{-\infty}^{+\infty} d\varepsilon \int_0^{2\pi} d\phi \, [f(\varepsilon) - f(\varepsilon + eV)] \text{Re}\left\{\frac{\varepsilon + eV + i\Gamma}{(\varepsilon + eV + i\Gamma)^2 - \Delta(\phi)^2}\right\} \quad (S1)$$

Here $f(\varepsilon)$ is the Fermi function containing the information of temperature, and $\Gamma$ is the scattering factor. For the spectrum measured on 1 QL Bi$_2$Te$_3$/FeTe$_{0.55}$Se$_{0.45}$ heterostructure, one gap is not enough for the fitting, so we use two gaps, i.e., a twofold SC gap and a fourfold SC gap, to fit the data. The differential conductivity in two-gap model is constructed as: $G = \alpha dI_1/dV + (1-\alpha)dI_2/dV$. Here $\alpha$ is the related spectral weight, and $I_{1(2)}(V)$ is the tunneling current contributed by different SC gaps. In addition, both $I_1(V)$ and $I_2(V)$ can be described by eq. S1. A twofold SC gap can be expressed as $\Delta = \Delta_{max}[1-\beta(1-\cos2\phi)]$, while a fourfold one is expressed as $\Delta = \Delta_{max}[1-\beta(1-\cos4\phi)]$. Here $\beta$ is the gap anisotropy. The SC functions and fitting parameters are shown in table S1.

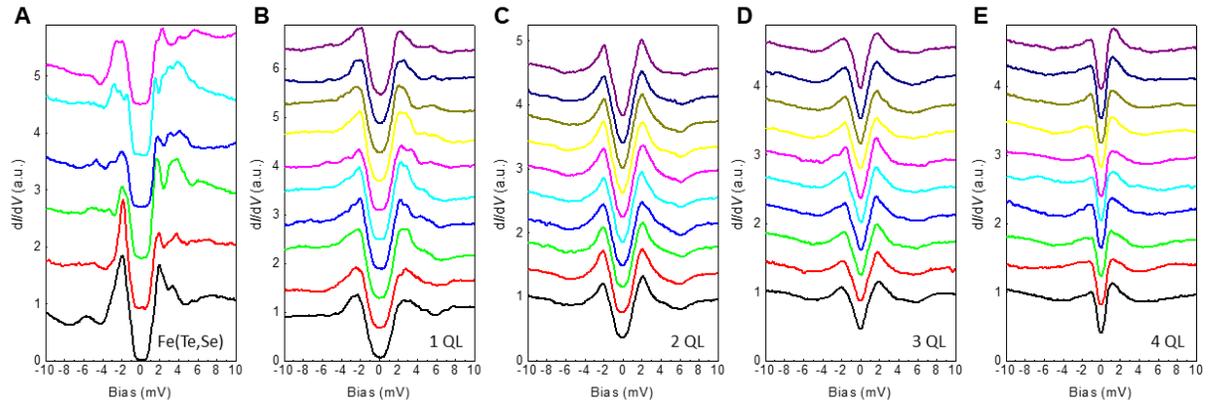

**Figure S1.** Series of tunneling spectra measured on FeTe$_{0.55}$Se$_{0.45}$ and Bi$_2$Te$_3$ thin films at 0.4 K and zero magnetic field. The data were taken along lines with lengths of 10, 27, 68, 68, and 39 nm, respectively. $V_{set} = 10$ mV, $I_{set} = 50$ pA.

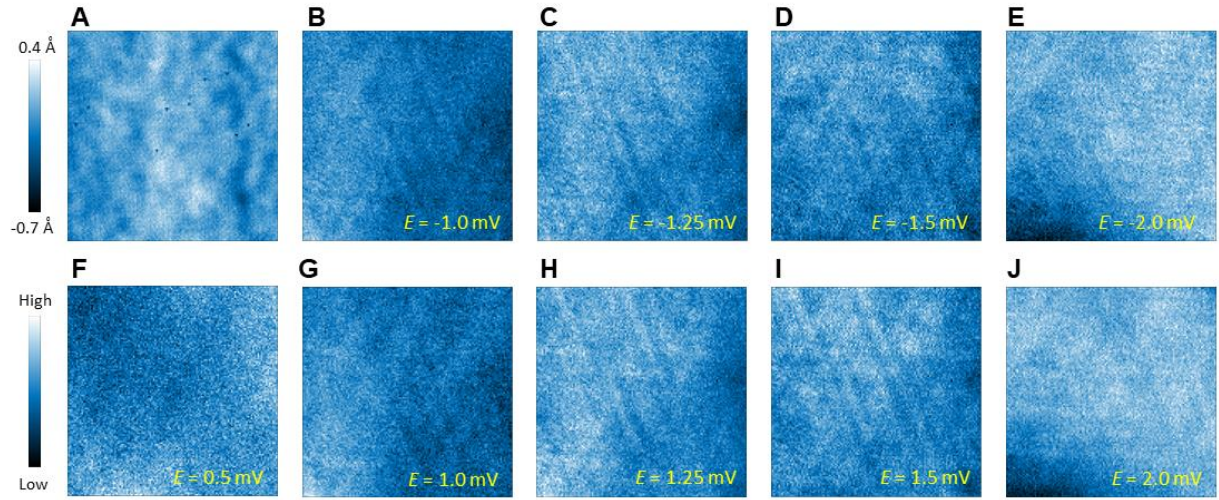

**Figure S2.** Real-space QPI patterns at series of energies on the 2 QL Bi$_2$Te$_3$/FeTe$_{0.55}$Se$_{0.45}$ heterostructure. **(A)** Topography of the view for QPI measurement. The size is 84 × 84 nm$^2$ with 128 × 128 pixels. **(B-J)** QPI image in real space measured at 0.4 K with $V_{set}$ = 10 mV and $I_{set}$ = 50 pA.

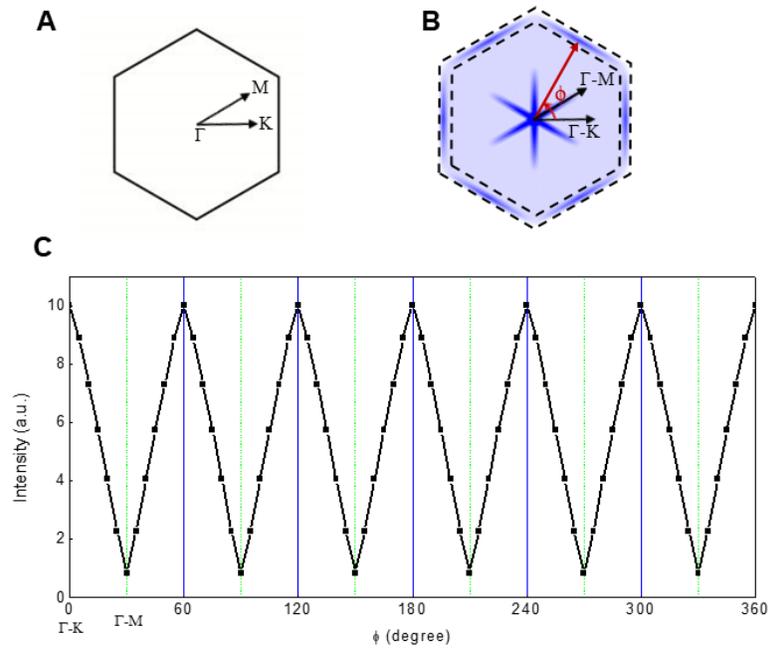

**Figure S3.** Simulation of FT-QPI pattern from a 2D hexagonal Fermi surface. (**A**) Schematic plot of 2D hexagonal Fermi surface with constant DOS around the Fermi surface. (**B**) Theoretical simulation of QPI scattering intensity by applying autocorrelation to **A**. (**C**) The averaged FT-QPI intensity per pixel in the *q*-space within the band bounded by the two parallel dashed hexagons in **B** with increment of every 5 degrees. The initial angle of $\phi$ is defined as one of the Γ-K directions.

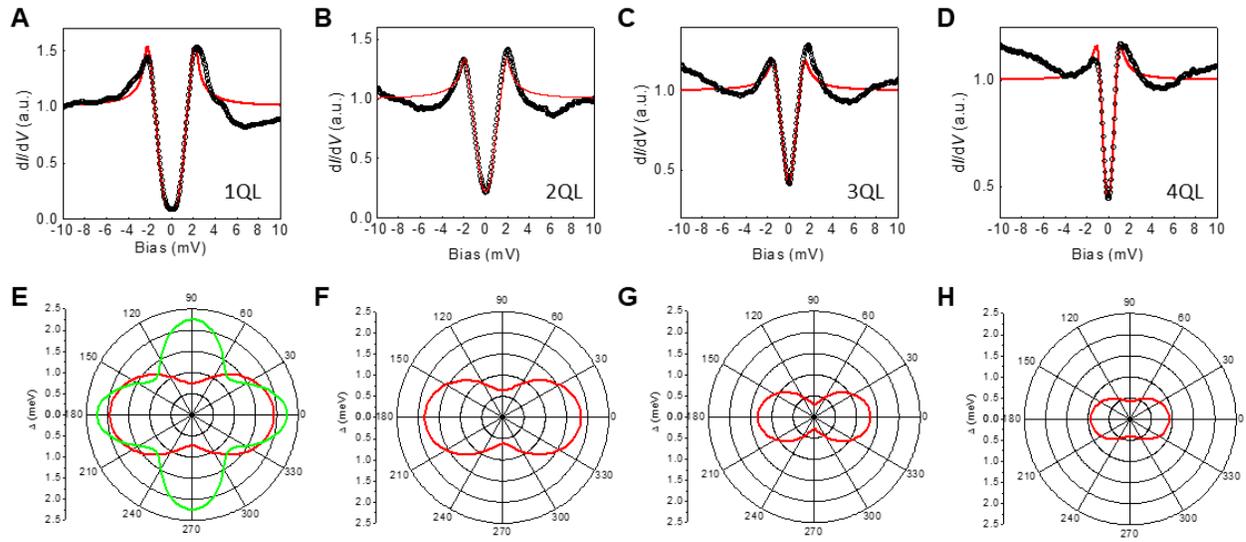

**Figure S4.** Typical tunneling spectra and theoretical fitting results. **(A to D)** Experiment data (solid symbols) and corresponding fitting curves (red lines) measured on 1 to 4 QL $Bi_2Te_3$/$FeTe_{0.55}Se_{0.45}$ heterostructure. **(E to H)** Gap functions used for fitting, respectively.

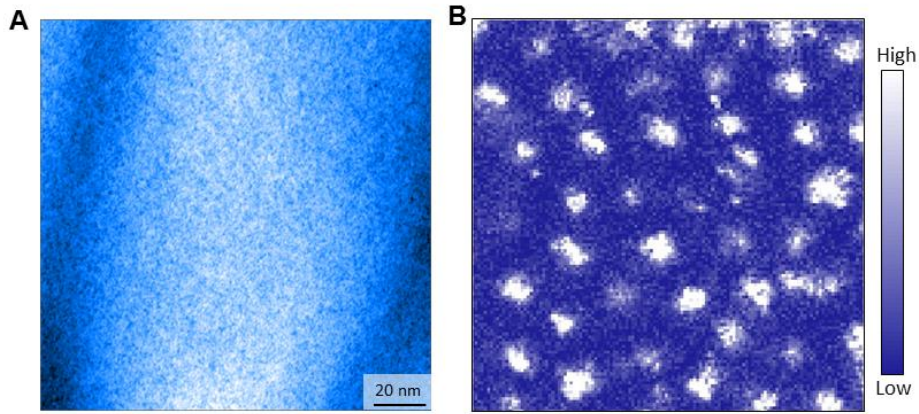

**Figure S5.** Vortex image on an area of FeTe$_{0.55}$Se$_{0.45}$ substrate without Bi$_2$Te$_3$ film. (**A**) Topography of FeTe$_{0.55}$Se$_{0.45}$ substrate, the size of the image is 150 × 150 nm$^2$. (**B**) Vortex image measured by zero-bias conductance map at 0.4 K and 0.7 T ($V_{set}$ = 10 mV, $I_{set}$ = 50 pA). The shapes of the vortices are usually round and isotropic.

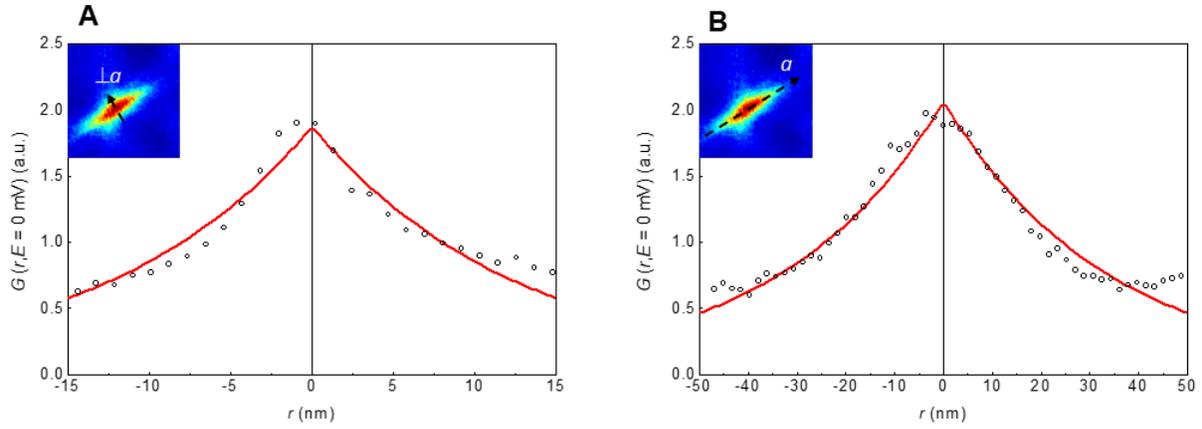

**Figure S6.** The spatial evolution of differential conductance across the elongated vortex. The experimental data are fitted by an exponential decay formula (red lines), and we can obtain different coherence length of $\xi_{\perp a}$ = 12.7 nm, $\xi_{\| a}$ = 33.5 nm.

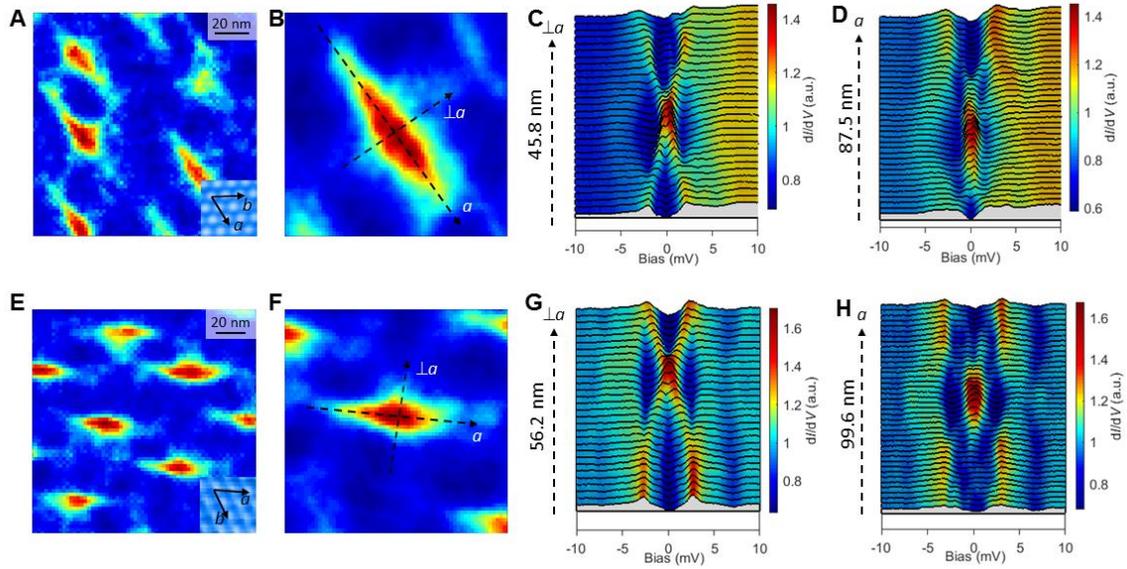

**Figure S7.** The control experiments of the elongated vortices. (**A** to **D**) Vortex images and bound states measured at different places on the same sample. (**E** to **H**) Vortex images and bound states measured on a different $Bi_2Te_3/FeTe_{0.55}Se_{0.45}$ heterostructure. All the vortices show elongated shape along one of the crystal axis. The spatial evolution of the vortex bound states look similar. Set point parameters $V_{set}$ = 10 mV, and $I_{set}$ = 50 pA.

**Table S1.** Superconducting gap functions and fitting parameters used in the fitting to the tunneling spectra measured on $Bi_2Te_3$ with different thickness at 0.4 K.

| Thickness | $\Gamma$ (meV) | $\Delta$ (meV) | $\alpha$ |
|---|---|---|---|
| 1QL | 0.11 | $\Delta_1 = 2.25(0.23 \cos4\phi+0.77)$ | 52% |
| | 0.11 | $\Delta_2 = 1.94(0.31 \cos2\phi+0.69)$ | 48% |
| 2QL | 0.25 | $\Delta = 1.85(0.33 \cos2\phi+0.67)$ | 100% |
| 3QL | 0.26 | $\Delta = 1.35(0.39 \cos2\phi+0.61)$ | 100% |
| 4QL | 0.31 | $\Delta = 0.93(0.28 \cos2\phi+0.72)$ | 100% |